\documentclass[aps,physrev,reprint,superscriptaddress,nobalancelastpage,citeautoscript,amsmath,amssymb,floatfix]{revtex4-1}
\usepackage{graphicx}
\usepackage{natbib}
\usepackage{dcolumn}
\usepackage{bm}
\usepackage[T1]{fontenc}
\usepackage{multirow}
\usepackage[%
  colorlinks=true,
  urlcolor=blue,
  linkcolor=blue,
  citecolor=blue
]{hyperref}

\newcommand{\tens}[1]{\overline{\overline{\bm{#1}}}}

\begin{document}

\title{Spin {H}all Magnetoresistance in Metallic Bilayers with In-plane Magnetized Ferromagnets}

\author{\L{}ukasz~Karwacki}
\email{karwacki@ifmpan.poznan.pl}
\affiliation{AGH University of Science and Technology, Department of Electronics, al. Mickiewicza 30, 30-059 Krak\'{o}w, Poland}
\affiliation{Institute of Molecular Physics, Polish Academy of Sciences, ul. M. Smoluchowskiego 17, 60-179 Pozna\'{n}, Poland}
\author{Krzysztof~Grochot}
\email{grochot@agh.edu.pl}
\affiliation{AGH University of Science and Technology, Department of Electronics, al. Mickiewicza 30, 30-059 Krak\'{o}w, Poland}
\affiliation{Faculty of Physics and Applied Computer Science, AGH University of Science and Technology, al. Mickiewicza 30, 30-059 Kraków, Poland}
\author{Stanis\l{}aw~\L{}azarski}
\affiliation{AGH University of Science and Technology, Department of Electronics, al. Mickiewicza 30, 30-059 Krak\'{o}w, Poland}
\author{Witold~Skowro\'{n}ski}
\affiliation{AGH University of Science and Technology, Department of Electronics, al. Mickiewicza 30, 30-059 Krak\'{o}w, Poland}
\author{Jaros\l{}aw~Kanak}
\affiliation{AGH University of Science and Technology, Department of Electronics, al. Mickiewicza 30, 30-059 Krak\'{o}w, Poland}
\author{Wies\l{}aw~Powro\'{z}nik}
\affiliation{AGH University of Science and Technology, Department of Electronics, al. Mickiewicza 30, 30-059 Krak\'{o}w, Poland}
\author{J\'{o}zef~Barna\'{s}}
\affiliation{Institute of Molecular Physics, Polish Academy of Sciences, ul. M. Smoluchowskiego 17, 60-179 Pozna\'{n}, Poland}
\affiliation{Faculty of Physics, Adam Mickiewicz University, ul. Uniwersytetu Pozna\'{n}skiego 2, 61-614 Pozna\'{n}, Poland}
\author{Feliks~Stobiecki}
\affiliation{Institute of Molecular Physics, Polish Academy of Sciences, ul. M. Smoluchowskiego 17, 60-179 Pozna\'{n}, Poland}
\author{Tomasz~Stobiecki}
\affiliation{AGH University of Science and Technology, Department of Electronics, al. Mickiewicza 30, 30-059 Krak\'{o}w, Poland}
\affiliation{Faculty of Physics and Applied Computer Science, AGH University of Science and Technology, al. Mickiewicza 30, 30-059 Kraków, Poland}

\begin{abstract}
We revisit the theory and experiment on spin Hall magnetoresistance (SMR) in bilayers consisting of a heavy metal (H) coupled to in-plane magnetized ferromagnetic metal (F), and determine contributions to the magnetoresistance due to SMR and anisotropic magnetoresistance (AMR) in four different bilayer systems: W/$\text{Co}_{20}\text{Fe}_{60}\text{B}_{20}$, W/Co, $\text{Co}_{20}\text{Fe}_{60}\text{B}_{20}$/Pt, and Co/Pt. To do  this, the AMR is explicitly included in the diffusion transport equations in the ferromagnet. The results allow precise determination of different contributions to the magnetoresistance, which can play an important role in optimizing prospective magnetic stray field sensors. They also may be useful in the determination of spin transport properties of metallic magnetic heterostructures  in other experiments based on magnetoresistance measurements.
\end{abstract}

\maketitle

\section{Introduction}
\label{sec:intro}

Spin Hall magnetoresistance (SMR) is a phenomenon that consists in resistance dependence on the relative orientation of magnetization and spin accumulation at the interface of ferromagnet and strong spin-orbit material  (such as $5d$ metals~\cite{Althammer2013,Nakayama2013,Chen2013,*Chen2016,Kim2016,Choi2017,Kawaguchi2018}, topological insulators~\cite{Lv2018}, or some 2D systems~\cite{Narayanapillai2017}). In such transition metals as W and Pt, the spin accumulation results from spin current driven by the spin Hall effect (SHE)~\cite{Hirsch1999,Dyakonov2007,Huang2012,Sinova2015}. The spin current diffuses then into the ferromagnet or exerts a torque on the magnetization while  being backscattered. Due to the inverse spin Hall effect (ISHE), the backscattered spin current is converted into a charge current that flows parallel to the bare charge current driven by external electric field, which effectively reduces the resistance~\cite{Chen2013,*Chen2016}. One of the most important advantages of driving spin currents by SHE is that the spin currents can be induced by a charge current flowing in the plane of the sample~\cite{Liu2019}. This may remedy some obstacles on the road to further miniaturization of prospective electronic components, which have been  encountered in spin-valves and magnetic tunnel junctions when the electric field is applied perpendicularly to interfaces. One of the drawbacks, however, is that the strength and effectiveness of such subtle effects depend strongly on the quality and spin properties of interfaces~\cite{Kobs2011,Pai2015,Zhang2015NPhys,Zhang2015,Tokac2015,Tokac2015PRL,Zhu2018,Amin2018}.

Although early SMR experiments were performed on heavy-metal/ferromagnetic-insulator bilayers~\cite{Althammer2013}, recent efforts are focused on the bilayers  with ferromagnetic metallic layers, such as Co or Co$_{20}$Fe$_{60}$B$_{20}$ ones~\cite{Kim2016,Kawaguchi2018},  which are currently more relevant for applications. When the magnetization is parallel to the spin accumulation,  the spin current from the heavy-metal can easily diffuse into the ferromagnetic metal (influencing its spin transport properties and spin accumulation on the ferromagnetic metal side)~\cite{Kim2016,Avci2015,Avci2015APL,Zhang2016,Avci2018,Taniguchi2015,Taniguchi2016,Yang2018}. This  is especially important when an additional spin sink (another  heavy-metal layer or an antiferromagnet) is on the other side of the ferromagnetic layer, where effects such as spin current interference might take place~\cite{Choi2017}.

Moreover, as charge current flows in plane of the sample, additional phenomena may occur, such as anisotropic magnetoresistance (AMR) or anomalous Hall effect (AHE)~\cite{Jan1957,McGuire1975,Bass2007,Bechthold2009AMR,Nagaosa2010,Iihama2018}. These effects can  obscure determination of spin transport parameters and make evaluation of the SMR contribution to the measured magnetoresistance more difficult. Since the determination of such transport properties as the spin Hall angle (which parameterizes strength of the spin Hall effect) and spin diffusion length in different experimental schemes, for instance in spin-orbit torque ferromagnetic resonance (SOT-FMR)~\cite{Liu2012,Skowronski2019},  relies heavily on the magnetoresistive properties of a system, it is important to properly determine all the contributions to magnetoresistance.

Here, we revisit the theory of spin Hall magnetoresistance in metallic bilayers by explicitly including the contributions from AMR and AHE into the spin drift-diffusion theory for the ferromagnetic metal layer. The expressions for magnetoresistance are then fitted to the data obtained from resistance meansurements on heavy-metal (H)/ferromagnet (F) bilayers, where H: W, Pt, while  F: Co, Co$_{20}$Fe$_{60}$B$_{20}$. This allows us to determine more accurately contributions from various magnetotransport phenomena occuring in metallic bilayers where the spin Hall effect is the driving source. Such analysis may also be useful in the efforts to optimize prospective devices for information technology.

The paper is organized as follows: Sec.~\ref{sec:theory} contains theoretical derivation of the formulas for multilayer magnetoresistance with explicit AMR and AHE contributions from metallic ferromagnet and SMR from H/F interface. Section~\ref{sec:exp} contains experimental details on the resistance and resistivity of the samples studied in this paper. Section~\ref{sec:results} contains results and discussion of the to experimental data on magnetoresistance. Finally, in Sec.~\ref{sec:summary} we briefly summarize the paper.

\section{Theory}
\label{sec:theory}

\begin{figure}
    \centering
\includegraphics[width=0.8\columnwidth]{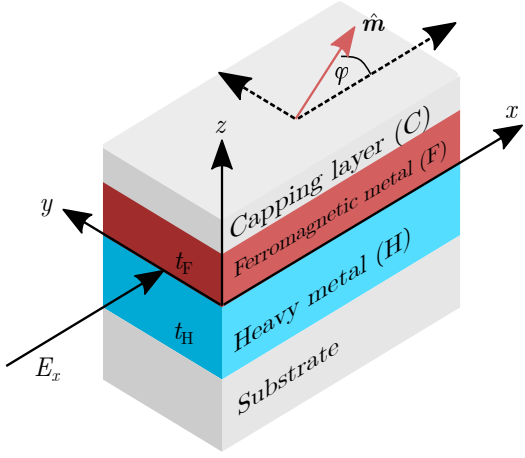}
\caption{Schematic representation of the system considered in this paper. The system consists of heavy-metal  (H) and ferromagnetic-metal (F) layers. The bilayer is deposited on a substrate and capped with a non-conductive layer (C). Thicknesses $t_H$ and $t_F$ of the H and F layers are varied, in general. Electric field, $E_x$, is applied along the $x$ axis, while the unit vector $\hat{\bm{m}}$ along magnetization of the ferromagnet is oriented in the plane and  at an angle $\varphi$ to the $x$ axis.}
\label{fig:model}
\end{figure}

Figure~\ref{fig:model} shows schematic representation of the system examined here. Note that the electric field in this figure is oriented along the axis $x$, but in order to properly capture the {\it planar Hall effect}  in calculations we consider a more general orientation of the field,
\begin{equation}
\bm{E}=E_x\hat{\bm{e}}_x+E_y\hat{\bm{e}}_y ,
\end{equation}
where $\hat{\bm{e}}_x$ and $\hat{\bm{e}}_y$ are unit vectors along the axes $x$ and $y$, respectively.

The spin current density tensor (measured in the units of charge current density)  in the heavy-metal (H) layer can be written as follows:
\begin{align}
    \frac{2e}{\hbar}\overline{\overline{\bm{q}}}_s^{\text{H}}=\sum_i\hat{\bm{e}}_i\bm{j}_{si}^{\text{H}}\,,
\end{align}
where $i=x,y,z$, and in the r.h.s. we used the convention for dyadic vector products according to which the first vector denotes the flow direction, and the second vector describes the  spin polarization and magnitude of the spin current.
Only the component flowing along the normal to interfaces is relevant and will be taken into account in the following, i.e. $\hat{\bm{e}}_z\cdot\frac{2e}{\hbar}\overline{\overline{\bm{q}}}_s^{\text{H}}\equiv\bm{j}_s^{\text{H}}(z)$, where
\begin{align}
\bm{j}_s^{\text{H}}(z)=-\frac{\theta_{\text{SH}}}{\rho_0^{\text{H}}}\hat{\bm{e}}_z\times\bm{E}+\frac{1}{2e\rho_0^{\text{H}}}\frac{\partial \boldsymbol{\mu}_s^{\text{H}}(z)}{\partial z}\,.
\end{align}
Here $\theta_{\text{SH}}$ is the spin Hall angle, $\rho_0^{\text{H}}$ is the bare resistivity of the heavy metal, and  $\boldsymbol{\mu}_s^{\text{H}}(z)$ is the spin accumulation that is generally $z$-dependent.

The charge current density in the heavy-metal (H) layer, in turn, can be written in the form
\begin{equation}
\bm{j}_c^{\text{H}}(z)=\frac{1}{\rho_0^{\text{H}}}\bm{E}+\frac{\theta_{\text{SH}}}{2e\rho_0^{\text{H}}}\hat{\bm{e}}_z\times\frac{\partial \boldsymbol{\mu}_s^{\text{H}}(z)}{\partial z}\,,
\end{equation}
and contains the bare charge current density and the current due to inverse spin Hall effect. Note, that the  spin current in general can induce charge current also flowing along the axes $x$ and $y$. However, due to lateral dimensions of the samples much larger than the layer thicknesses and spin diffusion lengths, those additional components can be neglected.

Analogously to the above, we define the spin current density tensor (in the units of charge current density) in the ferromagnetic layer (F)  as
\begin{align}
\frac{2e}{\hbar}\tens{q}_s^{\text{F}}=\bm{j}_s^{\text{F}}\hat{\bm{m}}\,.
\end{align}
However, now the first vector on r.h.s. describes direction flow and magnitude of spin current while the second one spin polarization, which now is along the magnetization. Thus, one can write $\frac{2e}{\hbar}\tens{q}_s^{\text{F}}\cdot\hat{\bm{m}}=\bm{j}_s^{\text{F}}$, where $\bm{j}_s^{\text{F}}$ is given by the equation ~\cite{Taniguchi2015,Taniguchi2016}:
\begin{align}
\bm{j}_s^{F}(z)&=
\frac{1}{2e\rho_0^{\text{F}}}\nabla\mu_s^{\text{F}}(z)+\frac{\beta}{2e\rho_0^{\text{F}}}\nabla\mu_c^{\text{F}}(\bm{r})\nonumber \\
&+\frac{\theta_{\text{AH}}}{2e\rho_0^{\text{F}}}\hat{\bm{m}}\times\nabla\mu_s^{\text{F}}(z)-\frac{\theta_{\text{AMR}}}{2e\rho_0^{\text{F}}}\hat{\bm{m}}\left[\hat{\bm{m}}\cdot\nabla\mu_s^{\text{F}}(z) \right]\,,
\end{align}
in which $\theta_{\text{AH}}$ and $\theta_{\text{AMR}}$ are the anomalous Hall angle and AMR angle, defined as $\theta_{\text{AH}}=\sigma_{\text{AH}}\rho_0^{\text{F}}$ and $\theta_{\text{AMR}}=\sigma_{\text{AMR}}\rho_0^{\text{F}}$, respectively, while  ${\mu_c^{\text{F}}(\bm{r})=2e\bm{E}\cdot\bm{r}+\mu_c^{F}(z)}$ is the electrochemical potential.

Charge current density in the ferromagnetic layer (F) can be written as~\cite{Taniguchi2015,Taniguchi2016},
\begin{align}
\bm{j}_c^{F}(z)&=\frac{1}{2e\rho_0^{\text{F}}}\nabla\mu_c^{\text{F}}(\bm{r})+\frac{\beta}{2e\rho_0^{\text{F}}}\nabla\mu_s^{\text{F}}(z)\nonumber \\
&+\frac{\theta_{\text{AH}}}{2e\rho_0^{\text{F}}}\hat{\bm{m}}\times\nabla\mu_c^{\text{F}}(\bm{r})-\frac{\theta_{\text{AMR}}}{2e\rho_0^{\text{F}}}\hat{\bm{m}}\left[\hat{\bm{m}}\cdot\nabla\mu_c^{\text{F}}(\bm{r}) \right].
\end{align}
Note, in the above equations for the current densities in both H and F layers  we assumed linear response to electric field, i.e. we neglected the so-called unidirectional spin Hall magnetoresistance effect~\cite{Avci2015,Avci2015APL,Zhang2016,Avci2018}.

The interfacial spin current density tensor can be written as
\begin{align}
\frac{2e}{\hbar}\tens{q}_s^{\text{HF}}=\hat{\bm{e}}_z\bm{j}_s^{\text{HF}}\,,
\end{align}
where the spin current $\bm{j}_s^{\text{HF}}$ flowing through the heavy-metal/ferromagnet interface is given by the following expression~\cite{Brataas2006}:
\begin{align}
\bm{j}_s^{\text{HF}}&=G_F\bigg[\left(\boldsymbol{\mu}_s^{\text{F}}-\boldsymbol{\mu}_s^{\text{H}}\right)\cdot\hat{\bm{m}}\bigg]\hat{\bm{m}}+G_i\hat{\bm{m}}\times\boldsymbol{\mu}_s^{\text{H}}(0)\nonumber \\
&+G_r\hat{\bm{m}}\times\hat{\bm{m}}\times\boldsymbol{\mu}_s^{\text{H}}(0)\,.
\end{align}
Here $G_F=(1-\gamma^2)(G_{\uparrow}+G_{\downarrow})/2$ with $\gamma$ defined as $\gamma=(G_\uparrow-G_\downarrow)/(G_{\uparrow}+G_{\downarrow})$ and $G_\uparrow$ and $G_\downarrow$ denoting the interface conductance for spin-$\uparrow$ and spin-$\downarrow$. Furthermore,  $G_{r}\equiv\operatorname{Re}G_{\text{mix}}$  and  $G_{i}\equiv\operatorname{Im}G_{\text{mix}}$, where $G_{\text{mix}}$ is the so-called spin-mixing conductance. Note, that we neglect explicitly a contribution from the interfacial Rashba-Edelstein spin polarization~\cite{Skowronski2019}.
A strong interfacial spin-orbit contribution which induces spin-flip processes can also be combined with the interfacial spin conductance $G_F$ as a spin-conductance reducing parameter $1-\eta$, with $\eta=0$ for no interfacial spin-orbit coupling, and $\eta=1$ for maximal spin-orbit coupling. Note, that this reduction could also be attributed to the magnetic proximity effect, especially in the case of Pt-based heterostructures~\cite{Huang2012}, however recent studies suggest its irrelevance for spin-orbit-torque--related experiments~\cite{Zhu2018}. In the following discussion we assume $\eta=0$ and treat $G_F$ as an effective parameter.

To find charge and spin currents we need  to find first the spin accumulation at the H/F interface and also at external surface/interfaces. This can be found from the following boundary conditions:
    \begin{subequations}
\begin{gather}
\bm{j}_s^{\text{H}}(z=-t_H)=0 \,, \\
\bm{j}_s^{\text{F}}(z=t_H+t_F)=0 \,, \\
\bm{j}_s^{\text{H}}(z=0)=\bm{j}_s^{\text{HF}}\,, \\
j_{s,z}^{\text{F}}(z=0)=\bm{j}_s^{\text{HF}}\cdot\hat{\bm{m}}\,.
\end{gather}
\end{subequations}
Having found spin accumulation and also electrochemical potential, one can find  the longitudinal ($l$) and transversal ($t$) in-plane components of the averaged charge current  $\bm{j}(\hat{\bm{m}})$ from the formula:
\begin{eqnarray}
j_{l(t)}(\hat{\bm{m}})=\frac{1}{t_H+t_F}\left[\int_{t_H}dz \hat{\bm{e}}_{x(y)}\cdot\bm{j}_c^{\text{H}}(z)\right.\nonumber \\
\left. +\int_{t_F}dz \hat{\bm{e}}_{x(y)}\cdot\bm{j}_c^{\text{F}}(z)\right]\,.
\end{eqnarray}
The total charge current can be written down in the Ohm's-law form,
\begin{align}
\bm{j}(\hat{\bm{m}})=\overline{\overline{\bm{\sigma}}}(\hat{\bm{m}})\bm{E}\,,
\end{align}
where the conductivity (resistivity) tensor takes the form:
\begin{align}
    \label{eq:sigtensor}
    \tens{\sigma}(\hat{\bm{m}})&\equiv\left[\tens{\rho}(\hat{\bm{m}})\right]^{-1}\nonumber \\
    &=
    \begin{bmatrix}
        \sigma_0+\sigma_xm_x^2+\sigma_ym_y^2 & m_xm_y\sigma_{xy} \\
        m_xm_y\sigma_{xy} & \sigma_0+\sigma_ym_x^2+\sigma_xm_y^2 \\
    \end{bmatrix}\,,
\end{align}
with
\begin{align}
\label{eq:sigcoeff0}
\sigma_0&=-2 \frac{\theta_{\text{SH}}^2}{\rho_0^{\text{H}}}\frac{\lambda_{\text{H}}}{t_F+t_H}\tanh \left(\frac{t_H}{2 \lambda_{\text{H}}}\right)+\frac{\rho_0^{\text{F}} t_H+\rho_0^{\text{H}} t_F}{\rho_0^{\text{F}} \rho_0^{\text{H}} t_F+\rho_0^{\text{F}} \rho_0^{\text{H}} t_H} \nonumber \\
&\approx \frac{\rho_0^{\text{F}} t_H+\rho_0^{\text{H}} t_F}{\rho_0^{\text{F}} \rho_0^{\text{H}} t_F+\rho_0^{\text{F}} \rho_0^{\text{H}} t_H} \\
\sigma_x&=\frac{\theta_{\text{SH}}^2}{\rho_0^{\text{H}}}\frac{g_{\text{DL}}\lambda_{\text{H}} }{t_F+t_H}\tanh \left(\frac{t_H}{2 \lambda_{\text{H}}}\right)-\frac{\theta_{\text{AMR}} }{\rho_0^{\text{F}}}\frac{t_F}{t_F+t_H}\,, \nonumber \\
&=\sigma_x^{\text{SH}}+\sigma_x^{\text{AMR}}\,,\\
\label{eq:sigcoeffw}
\sigma_y&=\frac{\theta_{\text{SH}}^2}{\rho_0^{\text{H}}}\frac{
g_H^{\text{F}}\lambda_{\text{H}}}{t_F+t_H}\tanh\left(\frac{t_H}{2 \lambda_{\text{H}}}\right)\nonumber \\
&-\frac{\theta_{\text{AH}}^2}{\rho_0^{\text{F}}} \frac{g_F^{\text{H}}\lambda_{\text{F}}}{t_F+t_H}\frac{\beta ^2}{\left(1-\beta ^2\right)}\tanh\left(\frac{t_F}{2 \lambda_{\text{F}}}\right)\nonumber \\
   &+2\frac{\theta_{\text{AH}}^2}{\rho_0^{\text{F}}}\frac{\lambda_{\text{F}}}{t_F+t_H}\frac{\beta ^2}{\left(1-\beta ^2\right)}   \tanh \left(\frac{t_F}{2 \lambda_{\text{F}}}\right)\nonumber \\
   &+\frac{\theta_{\text{AH}}^2}{\rho_0^{\text{F}}}\frac{t_F}{t_F+t_H} \nonumber \\
   &=\sigma_y^{\text{SH}}+\sigma_y^{\text{AH}}\,, \\
\sigma_{xy}&=\sigma_x+\sigma_y\,.
\end{align}
In the above expressions the following dimensionless  coefficients have been introduced to simplify the notation:
\begin{align}
    g_{\text{DL}}&=\left[1-\operatorname{sech}{\left(\frac{t_H}{\lambda_{\text{H}}}\right)}\right]\frac{g_r(1+g_r)+g_i^2}{(1+g_r)^2+g_i^2} \,, \nonumber \\
    g_{r,i}&=2G_{r,i}\rho_0^{\text{H}}\lambda_{\text{H}}\coth{\left(\frac{t_H}{\lambda_{\text{H}}} \right)}\,, \nonumber \\
       g_{H}^{\text{F}}&=\left[1-\operatorname{sech}{\left(\frac{t_H}{\lambda_{\text{H}}}\right)}\right]\frac{1}{1+\left(\frac{1}{2G_F\lambda_{\text{H}} \rho_0^{\text{H}}}+\gamma_H^{\text{F}}\right)\tanh \left(\frac{t_H}{\lambda_{\text{H}}}\right)}\,, \nonumber \\
       g_F^{\text{H}}&=\left[1-\operatorname{sech}{\left(\frac{t_F}{\lambda_{\text{F}}}\right)}\right]\frac{1}{1+\left(\frac{1}{2 G_F\lambda_{\text{F}} \rho_0^{\text{F}}}+\gamma_F^{\text{H}}\right)\tanh \left(\frac{t_F}{\lambda_{\text{F}}}\right)}\,, \\
       \gamma_{H}^{\text{F}}&=\frac{\lambda_{\text{F}} \rho_0^{\text{F}}}{\lambda_{\text{H}} \rho_0^{\text{H}}\left(1-\beta ^2\right)}\coth \left(\frac{t_F}{\lambda_{\text{F}}}\right)\,,\nonumber \\
       \gamma_F^{\text{H}}&=\frac{\lambda_{\text{H}}
          \rho_0^{\text{H}}\left(1-\beta ^2\right)}{\lambda_{\text{F}} \rho_0^{\text{F}} } \coth \left(\frac{t_H}{\lambda_{\text{H}}}\right)\,.
\end{align}
With the resistivity defined in Eq.~(\ref{eq:sigtensor}) we can now define magnetoresistance,
\begin{align}
\label{eq:MR}
MR=\frac{\rho_{xx}(\hat{\bm{m}}\parallel \hat{\bm{e}}_x)-\rho_{xx}(\hat{\bm{m}}\parallel \hat{\bm{e}}_y)}{\rho_{xx}(\hat{\bm{m}}\parallel \hat{\bm{e}}_x)}\,,
\end{align}
where $\rho_{xx}(\hat{\bm{m}})\equiv\overline{\overline{\bm{\rho}}}(\hat{\bm{m}})\cdot\hat{\bm{e}}_x$. Taking into account Eqs.~(\ref{eq:sigcoeff0})-(\ref{eq:sigcoeffw}), the above formula can be written as,
\begin{align}
MR\approx\frac{\sigma_y-\sigma_x}{\sigma_0}\,.
\end{align}
In order to compare the models with and without AMR and AHE, we define SMR as:
\begin{equation}
SMR=-MR\Bigg|_{\substack{\theta_{\text{AMR}}\to 0 \\ \theta_{\text{AH}}\to 0}}\,,
\end{equation}
which simplifies our model to that introduced by Kim et. al~\cite{Kim2016}.

\section{Experiment}
\label{sec:exp}

\begin{figure}
    \centering
\includegraphics[width=0.99\columnwidth]{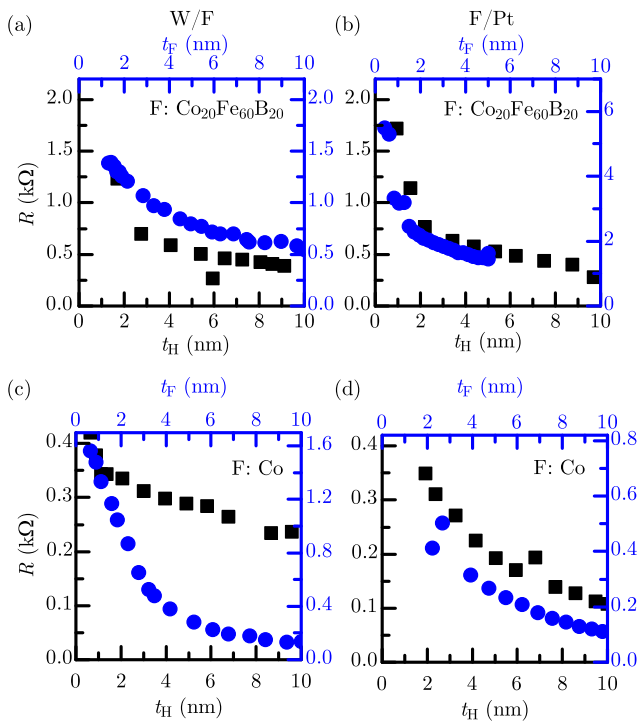}
\caption{Resistance $R$ as a function of thickness $t_{\text{H}}$ for indicated heavy metal layers (black squares): W [(a),(c)], Pt [(b),(d)], and as a function of thickness $t_{\text{F}}$ for ferromagnets (blue dots): $\text{Co}_{20}\text{Fe}_{60}\text{B}_{20}$ [(a),(b)] and Co [(c),(d)].}
\label{fig:RvsT}
\end{figure}

\newcommand{\minitab}[2][l]{\begin{tabular}{#1}#2\end{tabular}}
\begin{table}
\caption{\label{tab:res}%
Composition of samples, and resistivities of heavy metal and ferromagnetic layers. Numbers in parentheses denote thickness (in nm) of the corresponding layer.
}
\begin{ruledtabular}
\begin{tabular}{ c c c c }
No. &
Sample &
\minitab[c]{$\rho_0^{\text{H}}$\\ ($\mu\Omega\text{cm}$)} &
\minitab[c]{$\rho_0^{\text{F}}$\\ ($\mu\Omega\text{cm}$)} \\
 \colrule
W1 & W(5)/$\text{Co}_{20}\text{Fe}_{60}\text{B}_{20}$($t_F$)/Ta(1) & 185 & 144 \\
W2 & W($t_H$)/$\text{Co}_{20}\text{Fe}_{60}\text{B}_{20}$(5)/Ta(1) & 166 & 144 \\
W3 & W(5)/Co($t_F$)/Ta(1) & 120 & 22 \\
W4 & W($t_H$)/Co(5)/Ta(1) & 120 & 30 \\
P1 & $\text{Co}_{20}\text{Fe}_{60}\text{B}_{20}$($t_F$)/Pt(3) & 95 & 102 \\
P2 & $\text{Co}_{20}\text{Fe}_{60}\text{B}_{20}$(5)/Pt($t_H$) & 151 & 161 \\
P3 & Co($t_F$)/Pt(4) & 55 & 18 \\
P4 & Co(5)/Pt($t_H$) & 24 & 57 \\
\colrule
\end{tabular}
\end{ruledtabular}
\end{table}

Table~\ref{tab:res} shows the multilayer systems that were produced for SMR studies. The magnetron sputtering technique was used to deposit multilayers on the Si/SiO$_2$ thermally oxidized substrates. Thickness of wedged layers were precisely calibrated by X-ray reflectivity (XRR) measurements. The details of sputtering deposition parameters as well as structural phase analysis of highly resistive W and Pt layers can be found in our recent papers~\cite{Skowronski2019,Lazarski2019}. In turn, structure analysis of the Co crystal phases grown on disoriented $\beta$-W can be found  in the Supplemental Material~\cite{supplement}.

After deposition, multilayered systems were nanostructured using either electron-beam lithography or optical lithography, ion etching and lift-off. The result was a matrix of Hall bars and strip nanodevices for further electrical measurements. The sizes of produced structures were: 100~$\mu$m x 10~$\mu$m or 100~$\mu$m x 20~$\mu$m. In order to ensure good electrical contact with the Hall bars and strips, Al(20)/Au(30) contact pads with dimensions of 100~$\mu$m x 100~$\mu$m were produced. Appropriate placement of the pads allows rotation of the investigated sample and its examination at any angle with respect to the external magnetic field in a dedicated rotating probe station using a four-points probe. The constant magnetic field, controlled by a gaussmeter exceeded magnetization saturation in plane of the sample and the sample was rotated in an azimuthal plane from -120$^{\circ}$ to +100$^{\circ}$.

The resistance of the system was measured with a two- and four-point technique using Keithley 2400 sourcemeters and Agilent 34401A multimeter. As shown in Fig.~\ref{fig:RvsT}, resistances of bilayers with amorphous ferromagnet $\text{Co}_{20}\text{Fe}_{60}\text{B}_{20}$ are about one order higher than these with polycrystalline Co.
The same results were obtained using both techniques. The thickness-dependent resistivity of individual layers was determined by method described in Ref.~\cite{Kawaguchi2018}, and by a parallel resistors model. For more details on resistivity measurements we refer the reader to Supplemental Material~\cite{supplement}.

\section{Results and discussion}
\label{sec:results}

\begin{figure*}
    \centering
\includegraphics[width=\linewidth]{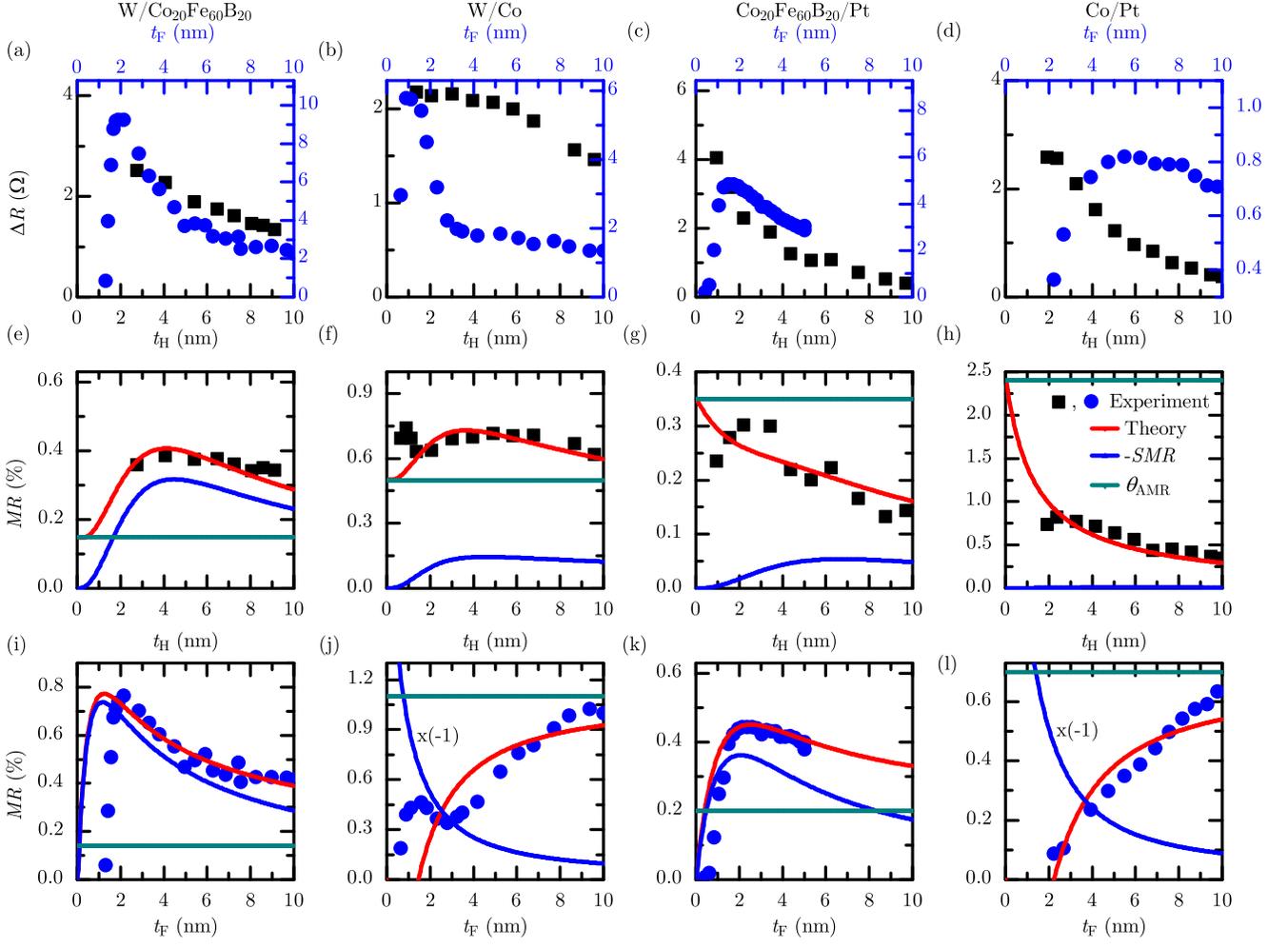}
\caption{Magnetoresistance $\Delta R$ as a function of both H (black squares) and F (blue dots) layers thicknesses [(a)-(d)]; relative magnetoresistance $MR$ for: W/$\text{Co}_{20}\text{Fe}_{60}\text{B}_{20}$ [(e),(i)], W/Co [(f),(j)], $\text{Co}_{20}\text{Fe}_{60}\text{B}_{20}$/Pt [(g),(k)], and Co/Pt [(h),(l)].}
\label{fig:MR}
\end{figure*}

\begin{table}
\caption{\label{tab:table1}%
Parameters used for fitting the model to experimental data for samples defined in Tab.~\ref{tab:res}.
}
\begin{ruledtabular}
\begin{tabular}{ c c c c c c c }
No. &
$|\theta_{\text{SH}}|$&
$\theta_{\text{AMR}}$ (\%)&
$G_r$ ($\Omega^{-1}\text{m}^{-2}$)&
$\lambda_{\text{H}}$ (nm)&
$\lambda_{\text{F}}$ (nm)&
 \\
 \colrule
W1 & 0.32 & 0.14 & $10^{13}$ & 1.3 & 1 \\
W2 & 0.28 & 0.13 & $10^{14}$ & 1.3 & 2 \\
W3 & 0.34 & 1.1 & $10^{17}$ & 1.3 & 5 \\
W4 & 0.47 & 0.5 & $10^{13}$ & 1.3 & 5 \\
P1 & 0.25 & 0.2 & $10^{11}$ & 2.2 & 2.5 \\
P2 & 0.09 & 0.35 & $10^{11}$ & 2.2 & 2.5 \\
P3 & 0.09 & 2.4 & $10^{13}$ & 2.2 & 5 \\
P4 & 0.28 & 0.7 & $10^{14}$ & 2.2 & 5 \\
\colrule
\end{tabular}
\end{ruledtabular}
\end{table}

 Table~\ref{tab:table1} shows parameters used for fitting the model to the experimental data on magnetoresistance. In order to simplify the analysis, we assumed thickness-independent interfacial spin-mixing conductances. However, it is well-known from the spin-pumping theory~\cite{Tserkovnyak2002} that the interfacial spin-mixing conductance is strongly dependent on Gilbert damping and saturation magnetization of the ferromagnet, which, in turn, depend on thickness of the layer. These parameters vary strongly mostly for thin layers, while they saturate for thicker ferromagnets. Moreover we assumed transparent contacts for parallel spin transport, i.e. $G_F\to\infty$, and also assumed $G_i$ to be neglible. Both assumptions are valid for metallic interfaces and the fitted parameters should be understood as upper limits. Although the theory predicts influence of AHE on the magnetoresistance, we neglect it in further discussion, as the so-called anomalous Hall angle is small in ferromagnets considered here, but should play an important role in out-of-plane magnetized systems or in-plane magnetized systems with more significant anomalous Hall angle. Finally, we assumed spin polarization $\beta=0.3$ for both Co and Co$_{20}$Fe$_{60}$B$_{20}$.

Figures~\ref{fig:MR}(a)--(d) show magnetoresistance as a function of both heavy metal and ferromagnetic metal layer thicknesses, while Figs.~\ref{fig:MR}(e)--(l) show relative magnetoresistance, on which we will focus our further discussion. As expected, SMR is larger in heterostructure with W as a heavy metal layer, than in the heterostructure with Pt, due to generally larger spin Hall angle of W, $-\theta_{\text{SH}}\approx\text{0.3 -- 0.4}$, compared to Pt, $\theta_{\text{SH}}\approx\text{0.1 -- 0.3}$. On the other hand, AMR is stronger in Co, $\theta_{\text{AMR}}\approx\text{0.5 -- 2.4\%}$, than in $\text{Co}_{20}\text{Fe}_{60}\text{B}_{20}$, for which $\theta_{\text{AMR}}\approx\text{0.13 -- 0.35\%}$.
Note, that we assume $\theta_{\text{AMR}}$ and $\theta_{\text{SH}}$ to be independent of layer thickness, which may result in overestimated parameters for very thin ferromagnetic layers.

Due to relatively high spin Hall angle in W, magnetoresistance as a function of heavy-metal layer thickness, shown in Figs.~\ref{fig:MR}(e) and~\ref{fig:MR}(f), has a large SMR component which is qualitatively and quantitatively similar for both W/$\text{Co}_{20}\text{Fe}_{60}\text{B}_{20}$ and W/Co bilayers. In W/Co, however, the total magnetoresistance is larger, due to larger AMR contribution of Co. In Pt bilayers, on the other hand, small spin Hall angle results in magnetoresistance dominated mostly by AMR contribution, as shown in Figs.~\ref{fig:MR}(g) and~\ref{fig:MR}(h).

The dependence of magnetoresistance on ferromagnetic layer thickness for fixed thickness of heavy-metal layers' is shown in Figs.~\ref{fig:MR}(i)--(l). For bilayers with $\text{Co}_{20}\text{Fe}_{60}\text{B}_{20}$, shown in Figs.~\ref{fig:MR}(i) and \ref{fig:MR}(k), AMR contribution is weaker than SMR for both W and Pt. In the case of Co-based heterostructures, shown in Figs.~\ref{fig:MR}(j) and \ref{fig:MR}(l), however, we observe negative SMR, which is dominated by parallel spin transport through the H/F interface. In this case large AMR might lead to strong overestimation of the relevant transport parameters. Moreover, the discrepancies between the model and the data can be attributed, as mentioned before, to varying spin-mixing conductance for small thickness of the ferromagnetic metal layer, possible magnetic dead layer, and magnetization that does not lie completely in-plane of the sample.

\section{Summary}
\label{sec:summary}

In conclusion, we have developed an extended model of magnetoresistance for magnetic metallic bilayers with in-plane magnetized ferromagnets, which explicitly treats both SMR and AMR contributions. The model was then fitted to experimental data on magnetoresistance in W/$\text{Co}_{20}\text{Fe}_{60}\text{B}_{20}$, W/Co, $\text{Co}_{20}\text{Fe}_{60}\text{B}_{20}$/Pt, and Co/Pt heterostructures to estimate the strength of SMR and AMR effects. These results allow for a more accurate estimation of different contributions to magnetoresistance in magnetic metallic systems, which is important for applications for instance in  spintronic memory read-heads or in other experimental schemes that rely on magnetoresistance measurements in evaluation of the spin transport properties.

\begin{acknowledgments}
This work is supported by the National Science Centre in Poland Grant UMO-2016/23/B/ST3/01430 (SPINORBITRONICS). WS acknowledges National Science Centre in Poland Grant No. 2015/17/D/ST3/00500. The nanofabrication process was performed at the Academic Centre for Materials and Nanotechnology (ACMiN) of AGH University of Science and Technology. We would like to thank M.~Schmidt and J.~Aleksiejew for technical support.
\end{acknowledgments}

\bibliography{karwacki_SMR.bib}

\end{document}